\documentclass[12pt]{article}
\usepackage{amsfonts}
\usepackage{amssymb}
\usepackage{amsmath}
\usepackage{graphicx}
\usepackage{psfrag}
\usepackage{epsfig}

\usepackage[normal,small]{caption}
\newcommand{\beq}{\begin{equation}}
\newcommand{\eeq}{\end{equation}}
\newcommand{\be}{\begin{equation}}
\newcommand{\ee}{\end{equation}}
\def\nref#1{(\ref{#1})}
\newcommand{\bea}{\begin{eqnarray}}
\newcommand{\eea}{\end{eqnarray}}

\def\Tr{{\rm Tr}}
\def\to{\rightarrow}

\def\m@th{\mathsurround=0pt }
\def\leftrightarrowfill{$\m@th \mathord\leftarrow
\mkern-6mu \cleaders\hbox{$\mkern-2mu \mathord- \mkern-2mu$}\hfill
 \mkern-6mu \mathord\rightarrow$}
\def\overleftrightarrow#1{\vbox{\ialign{##\crcr
     \leftrightarrowfill\crcr\noalign{\kern-1pt\nointerlineskip}
     $\hfil\displaystyle{#1}\hfil$\crcr}}}

\makeatletter
\renewcommand\section{\@startsection {section}{1}{\z@}%
                                   {-3.5ex \@plus -1ex \@minus -.2ex}%
                                   {2.3ex \@plus.2ex}%
                                   {\normalfont\large\bfseries}}

\renewcommand\subsection{\@startsection{subsection}{2}{\z@}%
                                   {-3.25ex\@plus -1ex \@minus -.2ex}%
                                   {1.5ex \@plus .2ex}%
       {\normalfont\normalsize\bfseries}}
\makeatother

\begin{document}
\begin{titlepage}
\begin{flushright}
UFIFT-HEP-06-5\\
PUPT-2193\\
\end{flushright}

\vskip 1.5cm

\begin{center}
\begin{Large}
{\bf Dynamics of Flux Tubes in Large $N$ Gauge Theories
}
\end{Large}

\vskip 1.5 cm

{\large I. R. Klebanov,\footnote{E-mail
address: {\tt klebanov@princeton.edu} } J. Maldacena,\footnote{E-mail
address: {\tt malda@ias.edu} } and C. B. Thorn\footnote{E-mail  address:
{\tt thorn@phys.ufl.edu}} }

\vskip 0.5cm
$^1${\it Joseph Henry Laboratories, Princeton University,
 Princeton, NJ 08544}

\vskip 0.5cm
$^2${\it School of Natural Sciences, Institute for Advanced Study,
Princeton, NJ 08540}

\vskip 0.5cm
$^3${\it Institute for Fundamental Theory\\
Department of Physics, University of Florida, Gainesville, FL
32611}

\vskip 1.0cm
\end{center}

\begin{abstract}
\noindent The gluonic field created by a static quark anti-quark pair is
described via the AdS/CFT correspondence by a string connecting
the pair which is
located on the boundary of AdS.
Thus
the gluonic field in a strongly coupled large $N$
CFT has a stringy spectrum of excitations. We trace the stability
of these excitations to a combination of large $N$ suppressions
and energy conservation.
Comparison of the physics of the $N=\infty$ flux tube in the
${\cal N}=4$ SYM theory at weak and strong coupling shows that
the excitations are present only above a certain critical
coupling.
The density of states
of a highly excited string with a fold reaching towards the
horizon of AdS is in exact agreement at strong coupling
with that of the near-threshold
states found in a ladder diagram model
of the weak-strong coupling transition.
We also study large distance
correlations of
local operators with a Wilson loop, and show that
the fall off at weak coupling
and $N=\infty$ (i.e. strictly planar diagrams)
matches the strong coupling predictions
given by the AdS/CFT correspondence, rather than those of
a weakly coupled $U(1)$ gauge theory.

\end{abstract}

\vfill
\end{titlepage}

\section{Introduction}

The usual picture of quark confinement is that the color flux
lines between the quark and antiquark do not spread throughout
space, as they do in electromagnetism, but rather form a narrow
tube.
Furthermore, this
flux tube (``QCD string'')
is supposed to have its own dynamics, since it can
support transverse oscillations. In fact, the hadron spectrum
consists of narrow resonances which may be associated with the
excitations of this flux tube. In the 't Hooft large $N$ limit
\cite{thooftlargen} the
self-interactions of the QCD string vanish, and we expect the gauge
theory to have a dual description in terms of
string theory with coupling $O( 1/N)$. The
string dynamics should
describe all excitations of the gauge theory flux tube. Since these
arguments are independent of whether the theory is confining or
not, we expect to have a string description even for non-confining
theories.

The AdS/CFT correspondence provides such a description; for example,
the string dual of the ${\cal N}=4 $ $SU(N)$
super Yang Mills theory is type IIB
string theory on $AdS_5\times S^5$ \cite{maldacena}. This theory is
conformal
and hence non-confining. Nevertheless,
the potential between a fixed quark and a fixed
anti-quark can be computed,
at infinite $N$ and large ${\hat\lambda}= g^2 N/4\pi^2$,
using the semiclassical string description
\cite{maldacenaqqbar}.
The resulting potential goes like
$-\sqrt{\hat\lambda}/L$, similar
to the weakly coupled gauge theory potential $-{\hat\lambda}\pi/L$.
Of course, this Coulombic form of the potential is a consequence
of the conformal symmetry. The fact that the $L$ dependence is the
same should not obscure the fact that the dynamics of the flux
tube at strong coupling is
dramatically
different than at weak coupling. In
fact, the string in $AdS_5$ has for large ${\hat\lambda}$
a rich spectrum of
discrete energy levels between $E\sim -\sqrt{\hat\lambda}/L$ and $E=0$,
reflecting the many degrees of freedom of a fluctuating
flux tube \cite{callang} (see also \cite{Bak}).
On the other hand, we will
show
that at weak coupling there is only an empty energy
gap between $E=-\pi{\hat\lambda}/L$ and $E=0$.
Thus there must be a critical coupling ${\hat\lambda}_c$ above which these
excitations begin to appear.

A simple way to see some indication of such a transition as a
function of the coupling is to consider the Klein-Gordon (or
Dirac) equation for a particle in the presence of a Coulomb
potential. These equations have a ``fall to the center''
instability when the coupling is larger than a critical value of
order one. The linearized  Yang-Mills  equations exhibit a similar
behavior \cite{mandula}
(see \cite{Shuryak} for a recent
discussion).

To explore the weak/strong coupling transition at large $N$,
we consider a soluble model
based on a truncated subset of planar Feynman diagrams, the ladder
graphs in Feynman gauge \cite{ericksonssz}. Although this model
gives an uncontrolled approximation, we will find that it captures
a surprising number of the expected features of the full answer.
For instance, there is a critical coupling where the number of
discrete states of the flux tube jumps from one to infinity. The
quark anti-quark ground state energy has the form $E_0 = - {
f(g^2N)/L}$. If we supply some energy $\Delta E > |E_0|$ then the
quarks can become unbound, which is to say that the system energy
becomes continuous. We will say in this situation that the system
ionizes and that $E=0$ is the ionization threshold. The rich
spectrum of bound states
mentioned above
arises when the total
energy of the system is less than zero. Of particular interest are
some discrete states with energies accumulating at the ionization
threshold. At strong coupling these states are described by folded
strings moving along the radial direction of $AdS_5$ with some
folds approaching the horizon. These states have similar
properties to the near threshold states appearing in the ladder
model above a critical coupling. In fact we will find a precise
matching between the density of states in the ladder approximation
and the density of such string states containing one fold. The
existence of this large number of near threshold states has
interesting consequences for the renormalization properties of
adjoint Wilson loops.

Another set of observables that shows an apparently new qualitative
behavior at strong coupling \cite{callang}
are the one point functions of local
operators in the presence of Wilson loops. In this case, however,
we will use
planarity and energetic considerations to show that the qualitative
behavior at large distances
is actually the same at weak and strong coupling.

\section{ Flux tube energy spectrum}

In this paper we consider mainly the ${\cal N}=4$ super Yang Mills
theory with fixed static quark and anti-quark sources.
They are represented
by two antiparallel time-like Wilson lines separated by a distance $L$.
We also have to
specify the locations on the $S^5$ for each of these lines,
which correspond to the orientations of their scalar charges. We
mainly choose
these locations to be the same, but will also comment on the situation
when they are different.

\subsection{Weak coupling analysis}

In Coulomb gauge, transverse gluons
decouple from the Wilson lines, so the lowest order system
energy is trivially the Coulomb potential
\bea
 E^{\rm singlet}_{0} &=&- { g^2 (N^2-1) \over 8 \pi N} { 1 \over L}\to
-{\pi{\hat\lambda}\over2L}
\label{bounds}\\
 E^{\rm adjoint}_{0} &=&+{ g^2  \over 8 \pi N} { 1 \over L}\to0
\eea
where the arrows show the $N\to\infty$ limits.
For the ${\cal N}=4$ case, $ E^{\rm singlet}_{0}=-\pi{\hat\lambda}/L$,
because of scalar exchange \cite{ericksonsz,grossd}.

At weak coupling, ${\hat\lambda}\ll1$,
we do not expect to find any other discrete states with
energy $ E^{\rm singlet}_0 < E <0$. One could imagine a state consisting of an
extra gluon that is interacting with the quark and antiquark
through planar interactions. However a simple argument
based on the uncertainty principle shows that such a
state has $E>0$. A massless particle localized within distance $r$
will have kinetic energy $\sim 1/r$ and potential energy $\sim - g^2 N/r$.
The kinetic energy is bigger at weak coupling, so we do not
expect a bound state.\footnote{One might wonder if there could be states
with energies $- (g^2 N)^n/L $, with $n>1$ which, at weak coupling could
be confused with the continuum.
However the argument based on the uncertainty
principle rules them out.}

However, there do exist states with unbound gluons alongside the
singlet $q{\bar q}$ system with continuous total energy $E\geq
E_0^{\rm singlet}$. This leads to the following interesting
effect. In analogy with QED we would have expected that if we put
the system in its lowest energy state (the singlet) and shake one
of the quarks a little bit, the system would decay immediately by
emitting massless radiation, since the theory has no gap. And,
indeed, this is what happens at finite $N$. However, if the
injected energy $\Delta E < |E^{\rm singlet}_0|$ the final state
of the quark anti-quark system must by energy conservation be the
color singlet ground state, which in turn implies that the emitted
gluons must be in an overall color singlet. But color singlet glue
emission is suppressed by powers of $1/N$, whether or not the
gluons are bound into glueballs. Thus we conclude that at weak
coupling and $N=\infty$, the energy spectrum of the quark
anti-quark system consists of a single discrete level $E^{\rm
singlet}_0$ separated by a gap from a continuous spectrum
$E\geq0$. At weak coupling we must think of the $N=\infty$ ``flux
tube'' connecting the quark to the antiquark as spread out all
through space (somewhat similar to the electric field of a dipole
in QED) but with a certain ``stiffness'' characterized by this
energy gap (quite unlike QED). But at weak coupling this stiffness
is too fragile to support vibrations. But we can imagine that as
the coupling increases and the gap widens, keeping $N=\infty$,
this stiffness becomes more robust eventually supporting
vibrations, populating the gap with discrete levels.

\subsection{Wilson loop at weak coupling}
In order to think in a clear way about the interpolation from weak to strong
coupling, it is useful to understand the results of the previous
subsection in terms of summing diagrams at large $N$. To
keep track of $1/N$ suppressions it is convenient to calculate
a single trace gauge singlet, and the Wilson loop $W(L,T)$
for an $L\times T$ rectangle fits the bill perfectly.
The energy spectrum of the flux tube can be read off from the
exponential time dependence of the expectation of the Wilson loop
$W(L,T)$ for an $L\times T$ rectangle.
\bea
\langle W(L,T)\rangle =\sum_i \rho_i e^{-TE_i}
\eea
or equivalently from the singularities in $E$ of its Laplace
transform, the resolvent
\bea
R(E)\equiv \int_0^\infty dT e^{ET}\langle W(L,T)\rangle
=\sum_i{\rho_i\over E_i-E}
\eea
Discrete eigenstates produce simple poles in $R$ and the continuum
produces cuts (implying imaginary parts) in $R$. The optical theorem
relates the imaginary part of $R$ to the decay rate of the system.

\begin{figure}[ht]
\psfrag{'a'}{${\hskip-2pt{}_{(a)}}$}
\psfrag{'b'}{${\hskip-2pt{}_{(b)}}$}
\psfrag{'c'}{${\hskip0pt{}_{(c)}}$}
\begin{center}
\includegraphics[height=1.5in,width=5in]{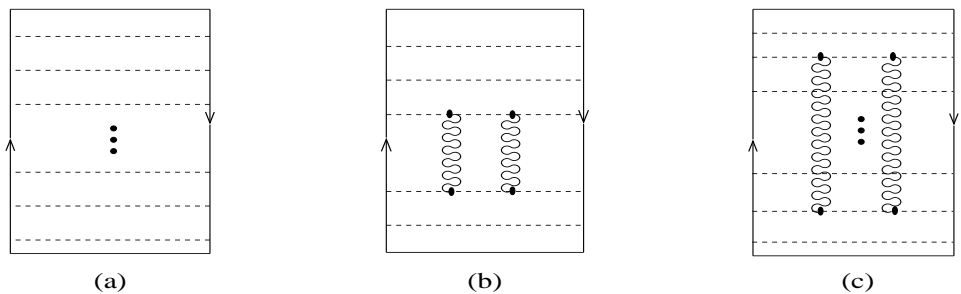}
\caption{Some Coulomb gauge Feynman diagrams contributing to the
Wilson loop. The wavy lines represent
transverse gluons and the dashed lines instantaneous Coulomb
exchange. The Coulomb exchange diagrams (a) sum up to the
ground state pole in $R(E)$. The planar diagram (b)
gives a contribution to the continuum
cut $E\geq0$. The gap destroying nonplanar diagram (c)
is suppressed at $N=\infty$.}
\label{H}
\end{center}
\end{figure}

Then the results of the previous subsection can be succinctly
summarized by writing, for $\hat \lambda \ll 1$,
 \bea
R(E)&\approx& {\rho_0\over -\pi{\hat\lambda}/L-E}+\int_0^\infty
dE^\prime {\rho(E^\prime) \over E^\prime - E}+{1\over
N^2}\int_{-\pi{\hat\lambda}/L}^\infty dE^\prime {\rho_1(E^\prime)
\over (E^\prime-E)} \eea where $\rho_0=O(1)$,
$\rho(E^\prime)=O(\hat \lambda)$, and $\rho_1(E^\prime)=O({\hat
\lambda}^2)$. The $N\to\infty$ limit removes the gap destroying
third term. We used the ground state energy of the ${\cal N}=4$
theory.  For pure Yang-Mills, just substitute ${\hat
\lambda}\to{\hat \lambda}/2$. In Fig.~\ref{H} we display a sample
Coulomb gauge Feynman diagram contributing to each of the terms.
\footnote{We chose diagrams with two transverse gluons because
their color states are more generic than a single gluon which only
exists in an adjoint representation.} By focusing on the infinite
$N$ Wilson loop, we cleanly remove the gap destroying processes
that are present at finite $N$.

The Wilson loop also has a transparent interpretation on the string side of the
duality, which makes
it a particularly apt observable for following
the weak/strong coupling interpolation.
In the next section we discuss the flux tube
spectrum at strong coupling. We will see that the main qualitative change
is the replacement of the single pole term with an infinite number of
pole terms accumulating at $E=0$.

\subsection{Strong coupling analysis}

At strong coupling we can study the system using the AdS/CFT
correspondence \cite{maldacena}. The ${\cal N}=4$ supersymmetric
$SU(N)$ gauge theory   is dual to type IIB superstring theory on
the manifold AdS$_5\times$S$_5$   whose metric is
\begin{equation} \label{adsmet}
ds^2 = {R^2\over z^2} (dz^2 + dx_\mu^2) + R^2 d\Omega_5^2 \ ,
\end{equation}
and $R^2 = \sqrt{g^2 N} \alpha'$. In the large $N $ limit the
superstrings   are noninteracting,
just as glueballs would be in a confining gauge theory.
Configurations with external quarks were studied in
\cite{maldacenaqqbar}.  The absence of confinement
implies that the energy of an isolated quark has no IR divergence.
The dual of a quark at spatial position $\vec x$ in the gauge
theory is a type IIB string stretched along the $z$ direction all
the way to infinite $z$ \cite{maldacenaqqbar}. The energy of such
a string is
\begin{equation} \label{potenq}
{R^2\over 2\pi\alpha'} \int_\epsilon^\infty {dz\over z^2} \ ,
\end{equation}
where we have introduced a UV cut-off at small $z=\epsilon$. The
IR corresponds to the large $z$ region and we see that
\nref{potenq} is finite there.

\begin{figure}[ht]
\begin{center}
\includegraphics{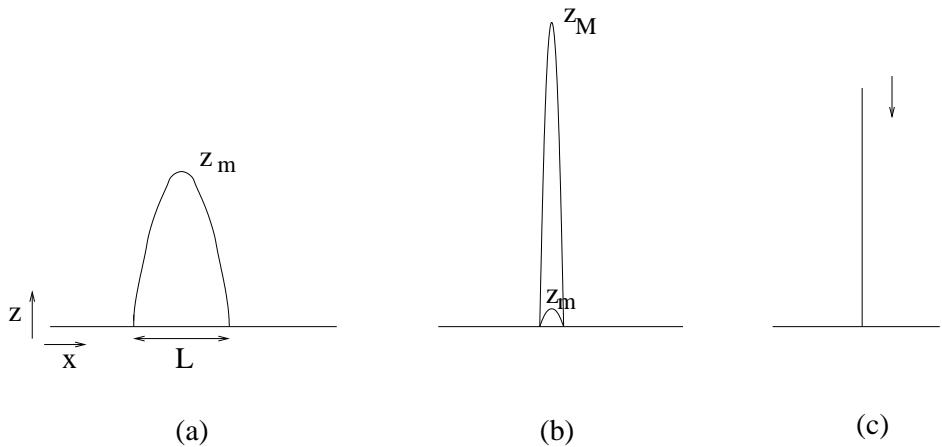}
\caption{In (a) we show the string going between two points in the
boundary which gives the dual description of the flux tube going
between the quark and the anti-quark. In (b) we stretch the string
up to a large $z_M \gg z_m$. In (c) we replace the stretched
string in (b) by a folded string sitting at a single spatial point
on the boundary.  } \label{stretched}
\end{center}
\end{figure}

If we consider a quark at $x_1=-L/2$ and an anti-quark at
$x_1=L/2$, whose colors are different, they cannot form a color
singlet bound state. The corresponding state in AdS space has two
straight strings, one running from $z=\epsilon$ to $\infty$ at
$x_1= -L/2$, and the other running from $z=\infty$ to $\epsilon$
at $x_1=L/2$. If the colors are correlated so that a $q\bar q$
colorless bound state can form, then the dual configuration is
given by a single segment of the string
(see figure \ref{stretched}(a)) whose position $z_c(x)$
was determined in \cite{maldacenaqqbar}; it satisfies
\begin{equation}
\left ( z_c' (x)\right )^2= {z_m^4\over  z_c^4 (x)}-1 \ .
\end{equation}
 The maximum value of the $z$-coordinate is attained
at the mid-point
\begin{equation}
z_m = z_c(0)= { \Gamma(1/4)^2L \over (2\pi)^{3/2} } \ .
\end{equation}
This string is the dual of the gluonic field that forms
between the static quark and anti-quark in the conformal gauge
theory.
The string energy has the separation dependence of a
Coulomb potential \cite{maldacenaqqbar}
\be \label{stingen}
E_0(L)=- c { \sqrt{g^2 N} \over L}
~,~~~~~~~~~~~c= { 4 \pi^2 \over \Gamma({1 \over 4})^4 }
\ee

Callan and Guijosa \cite{callang} studied
small oscillations about the static string connecting the heavy
quark and anti-quark (see also \cite{Bak}).
The equation for the oscillation modes transverse to the $z$ direction,
derived in \cite{callang}, may be
written as
\beq \label{transos}
\left [\partial_t^2 - {z_c^4 (x)\over z_m^4}
\partial_x^2 \right ] y(x,t) =0\ .
\eeq
After imposing the Dirichlet boundary conditions, $y(\pm L/2,
t)=0$, the exact normal mode frequencies are $\omega_n
=\xi_n/z_m$, with $\xi_n$ satisfying \bea \label{exactfreq}
\xi_n\sqrt{\xi_n^4-1}\int_0^1{t^2dt\over[1+\xi_n^2t^2]
\sqrt{1-t^4}}={n\pi\over2},\qquad n=1,2,\cdots \eea
This result follows from the quantization of
the phase in eq. (33) of \cite{callang}
and corrects a typo in eq. (38) of that reference.
The integral
in (\ref{exactfreq}) is elliptic, and the equation must be solved
numerically. However, for large frequency it becomes elementary
\cite{browertt}:
\bea \xi_n\sim
{(n+1)\pi\over2}\left[\int_0^1{dt\over\sqrt{1-t^4}}\right]^{-1}
=(n+1){({2\pi})^{3/2}\over\Gamma(1/4)^2},\qquad n \gg 1
\eea
This result is equivalent to the WKB approximation applied to
(\ref{transos}), but with $n$ replaced by $n+1$. The analysis of
small transverse oscillations thus gives discrete excitation
energy levels $E_{\{N_n\}}-E_0=\sum N_n\omega_n$. This stringy
spectrum is of the same qualitative nature as the spectrum around
a confining flux tube of length $L$, which can be deduced from the
Nambu-Goto action in the static gauge. Of course, in our
non-confining example, this small oscillation spectrum is only
valid if $E - E_0 \ll |E_0|$.
If ${\hat\lambda}$
is large but  finite we will need to go beyond the quadratic
approximation for computing the energy when
$\sum N_n \xi_n \sim \sqrt{g^2 N}$.

Now, let us also
derive the fluctuation equation for the
  ``longitudinal modes''
coming from the function $z(x,t)$.
In the
static gauge $\tau= t$, $\sigma= x$, the Nambu-Goto action is
\beq
S = -{R^2\over 2\pi \alpha'} \int dt \int_{-L/2}^{L/2} dx \ z^{-2}
\sqrt{ 1- (\dot z)^2 + (z')^2 } \ . \eeq
>From it we find the
equation of motion
\beq {\partial^2 z\over \partial
t^2} \left (1+ (z')^2 \right ) -{\partial^2 z\over \partial x^2}
\left (1- (\dot z)^2 \right )- {2\over z} \left ( 1- (\dot z)^2 +
(z')^2 \right ) - 2 {\partial^2 z\over \partial t \partial x}
{\dot z} z' = 0 \ . \eeq
Expanding around the minimal energy solution $z_c$ as
 $z = z_c(x) + u(x,t)/z_c(x)^2 $,
we find that the linearized equation for $u$ is
\beq \left [\partial_t^2 -
{z_c^4 (x)\over z_m^4}
\partial_x^2 \right ] u(x,t) = {2 z_c^2 (x)\over z_m^4} u(x,t)\ .
\label{longeq}
\eeq
Compared to the transverse
oscillation equation \nref{transos}
 there is an attractive potential
term on the right hand side.\footnote{This
equation is equivalent to the longitudinal
oscillation equation obtained using lightcone quantization
in \cite{browertt}.}
This equation has a series of eigenfrequencies $\tilde \omega_n=
\tilde\xi_n/z_m$, and each such oscillator can have a discrete
occupation number $\tilde N_n$. Thus, for the longitudinal modes
we again obtain a stringy form of the spectrum. Applying the WKB
approximation to (\ref{longeq}), gives
\bea \int_0^1\ dt{ \sqrt{\tilde\xi_n^2 + 2  t^2}
\over\sqrt{1-t^4}}= {n \pi\over2},\qquad n \gg 1
\eea

In order to analyze the solution with high occupation number we
resort to a different approximation. We consider the following
thought experiment. Let us stretch the string from $z=z_m$ to $z_M
\gg z_m$, see figure \ref{stretched}(b). Such a string will be near the
ionization threshold since $0 > E \gg E_0 $.
 In
this case, we can neglect the transverse separation $L \sim z_m
\ll z_M $ and consider the string as a folded string that lies at
one point in $x$ and extends purely along the $z$ direction, see figure
\ref{stretched}(c).  The
dynamics of these strings is very similar to the dynamics of
folded strings in two flat dimensions studied in \cite{folded}.
The spacetime picture for these folded strings is easy to
understand. We simply consider the tip of the
fold as a massless particle which is
being pulled by a string.
So we  consider
a folded string that is stretched along the $z$ direction and
is located at a point of the $R^3$ of the spatial coordinates of the
boundary. We put a UV cutoff in the radial direction at $z_m \sim L$.
We now stretch the string from $z_m$ to $z_M$, with $z_M \gg z_m$.
The energy of this stretched string is
\beq \label{energ}
 E = 2 { R^2\over 2\pi \alpha'}
 \int { dz \over z^2} = - 2  { R^2\over 2\pi \alpha'} { 1 \over z_M}
\ ,
\eeq
 where we subtracted a cut-off dependent constant in the energy in
 such a way that the energy becomes zero when $z_M \to \infty$.
We are interested in highly excited states with energies
  $ 0> E \gg E_0 \sim  - { 1 \over L}$.

 If we stretch the folded part of the string and then release it,  the tip of
 the string will move towards the boundary of $AdS$. The tip will be
 accelerated very quickly to nearly the speed of light. So, we can
 think of the tip of the string as a massless particle with
 momentum and energy $p^0 = |p_z|$. Energy conservation
implies
\beq \label{enerc}
 p^0 -  {R^2\over \pi \alpha'}   { 1 \over z}   = E~,~~~~~~~p^0 =
 |p_z| = {R^2\over \pi \alpha'} \left( { 1\over z} - { 1 \over z_M} \right)
\eeq
Using the Bohr-Sommerfeld quantization condition
\beq
  n = { 1 \over 2 \pi}  \oint p_z dz =
  {  R^2  \over \pi^2 \alpha' } \int_{z_m}^{z_M} dz
\left( { 1\over z} - { 1 \over z_M}\right) =
  {  R^2 \over \pi^2 \alpha' }\left[  \log { z_M \over z_m} - 1 + { z_m \over z_M
}  \right] \eeq
  To leading order we have $n \sim \log z_M/z_m $.
  Combining this with \nref{energ} we find that
\beq \label{enerfind}
  - \log (- E z_m)  \sim   { \pi^2 \alpha' \over  R^2  } \, n  \sim
      { \pi^2 \over \sqrt{g^2 N} } \,  n
 \eeq
 where we neglected a constant piece which could depend
on the details of the
 regularization procedure.
 We can think of \nref{enerfind} as
giving us a density of states $ { dn/d \log|E|}  $.
The conformal symmetry determines \nref{enerfind}
up to a function of the coupling. Thus, the
 AdS geometry was used to compute the square root appearing in \nref{enerfind}.
Note that this density of states is also related to the scattering
phase shift accumulated when the tip of the string moves from the
boundary to the region near the horizon and back.

\section{Comparison with the ladder appoximation}

By comparing the weak coupling and strong coupling results
it is clear that there
should exist a critical coupling where new states appear from the continuum.
To definitively analyze this issue would require a complete understanding
of the sum of all planar diagrams, a hugely ambitious program
beyond the scope of this paper.

In this section we will consider instead a simple, albeit uncontrolled,
approximation to the gauge theory computation: the sum of ladder
diagrams in Feynman gauge. This approximation was introduced in
\cite{ericksonssz} where it was shown that the
energy eigenvalue problem for the ladder model is equivalent to
the Schroedinger-like differential equation \bea \label{schr}
\left[-\partial^2_t-{\hat \lambda\over
L^2+t^2}\right]\psi=-{E^2\over4}\psi ~;~~~~~~~~~~~~ \hat \lambda
\equiv { g^2 N \over 4 \pi^2 } \eea where the energy $E$ is
defined so that threshold is at $E=0$. Later it was shown that the
analogous approximation for circular Wilson loops
\cite{ericksonsz,grossd}, the sum of ``rainbow'' graphs, is exact
for ${\cal N}=4$ at $N=\infty$, since all the more complicated
planar graphs apparently cancel out. It was even argued in
\cite{grossd} that it was exact to all orders in the $1/N$
expansion, see also \cite{othercircle}. The result for the energy
$E_0$ that we get from the ladder diagrams \nref{schr} does not
agree precisely with the the gravity answer but it has the same
dependence on the 't Hooft coupling \cite{ericksonssz}.

In fact this model shows a transition at $\hat\lambda_c=1/4$
between a single bound state and many bound states
\cite{browertt}. For very small $\hat \lambda$ we can approximate
the potential by a delta function \cite{ericksonssz}, which has
only one bound state and reproduces \nref{bounds}. On the other
hand, for very small energies $|E| \ll 1/L$ and for  $ L \ll |t|
\ll 1/|E|$ we can solve the equation \nref{schr} by setting $\psi
\sim |t|^{\alpha} $, with
 \be
  \alpha(\alpha-1) + \hat \lambda =0 ~,~~~~~~~\Rightarrow~~~ \alpha =
  { 1 \over 2} \pm i \sqrt{\hat \lambda - { 1\over 4} }
\ee
  We see that for $\hat \lambda > \hat \lambda_c = 1/4$
the wave function oscillates as
\be \label{cosf}
  \cos\left( i \left|\alpha-{1 \over 2} \right| \log |t|\right)
 ~ ,~~~~ {\rm for} ~~~  L \ll |t| \ll
1/|E|  \ . \ee
 We expect a turning point at around
$|t_{turn}| \sim 1/|E|$. Since the equation is
 scale invariant in this region, the details
of the turning point are   $E$ independent.
 By taking a small value of $E$ we can get
many zeroes of the cosine \nref{cosf} before
 getting to the turning point where \nref{cosf} ceases to be valid. This implies that
 we have an infinite number of bound states for $\lambda > \lambda_c$.
 Furthermore, the number of zeroes of \nref{cosf} between the turning points at $t = \pm
 t_{turn}$   gives us
 the number of states below the energy we are considering.
 Thus the number of states goes as
  \be
  n \sim 2 { \sqrt{\hat \lambda -{1\over 4} } \over \pi  }  \log |t_{turn}|  \sim
   - 2 { \sqrt{\hat \lambda - {1 \over 4} } \over \pi  } \log(-EL) \sim  -{ \sqrt{g^2 N }
   \over \pi^2 } \log(-E L)
 \ee
 where in the last relation we assumed $g^2 N \gg 1$.
This precisely reproduced the string theory answer \nref{enerfind}.
It would be interesting to find if there is a rationale
behind this agreement, since
the ladder model includes only a small subset of all planar diagrams.

On the other hand, for $\hat \lambda < \hat \lambda_c$,
there is only one bound state solution of
\nref{schr}.
To show this, we
set $E=0$ in \nref{schr} and solve the equation in terms
of hypergeometric functions. We then find that the solution with the right
boundary conditions at $t= -\infty$ has a single zero for $\hat \lambda < 1/4$
(but it does not obey the
right boundary conditions at $t=+\infty$).
Therefore, there is a single bound state for \nref{schr} in this
region\footnote{ The two solutions to \nref{schr} with $E=0$ are
$F[ \alpha_- , \alpha_+, {1 \over 2} ; - x^2 ], ~x F[ - \alpha_- ,
- \alpha_+, {3 \over 2} ; - x^2 ]$ where $ \alpha_\pm = - { 1\over
4} \pm { 1 \over 4} \sqrt{1 - 4 \hat \lambda} $. }.

Is there any sign, on the string theory side, of a transition
at a critical value of the radius?
A precise answer to this question would require a solution of
the string theory on $AdS_5 \times S^5$. We will content ourselves with the following
qualitative argument.
Let us consider again the folded string discussed above. Analyzing the string
theory more precisely, we find that the tip of the string is better modelled
by a massive particle. We can think of this massive particle
as a massive string state or a Kaluza Klein state on $AdS_5 \times S^5$.
So let us repeat the analysis of
\nref{enerc} for a massive particle at the tip of the string.
The action is ($T=1/(2\pi \alpha')$)
 \be \label{mends}
S = \int dt \left[ - m R {1 \over z} \sqrt{ 1- \dot z^2} + 2 R^2
T{ 1 \over z}
 \right]
 \ee
 Now we write the energy conservation condition
 \be\label{enerco}
 { m R \over z} { 1 \over \sqrt{ 1 - \dot z^2} } - 2 R^2 T { 1
 \over z} =  - 2R^2 T{ 1 \over z_M} + { m R \over z_M } = E\ ,
 \ee
which allows us to solve for the motion of the tip position $z$.
Let us imagine that the tip starts from rest at $z=z_M$.
For $m< 2RT$, the tip falls towards the boundary (small $z$).
For $z\ll z_M$, it reaches the
asymptotic velocity given by
 \be\label{asymptvel}
  \sqrt{ 1 - \dot z^2} = { m \over 2 T R}
\ .  \ee
For $m\ll 2TR$, the asymptotic speed
is very close to the speed of
  light.  This happens in the case that
$R \gg \sqrt{\alpha'}$, since $m$ is
  at most $1/\sqrt{\alpha'}$ for a folded string.
On the other hand, if the mass approaches $2TR$,
then the final velocity becomes non-relativistic,
  and the approximations we have made around \nref{enerc} are not
  valid. In fact,
 as we reduce the value of $R$ and imagine that
  the tip of the string stays with a mass of order $1/\sqrt{\alpha'}$,
   we see that negative
  energy solutions of \nref{enerco} cease to exist. In fact, for $m> 2TR$
the tip falls toward the horizon (large $z$), causing a
disintegration of the string. Hence, for $R\ll \sqrt{\alpha'}$ the
string cannot support any bound states. This provides a signal,
from the dual $AdS$ point of view, of a transition similar to the
transition at $\hat \lambda =1/4$ in the ladder diagrams. Of
course, this argument is only qualitative. We will have to wait
till the $AdS_5 \times S^5$ string theory is properly quantized in
order to study the transition on the $AdS$ side more precisely.

Another situation where this transition happens is the following.
 We now stay at strong coupling but consider different scalar
charges for the quark and antiquark. In that case the string ends
at two different points on $S^5$. In appendix A we display the
solution describing the asymptotic motion of such a string. In
order to get a qualitative idea we can model this situation by
assuming that the mass at the tip is of the order of $m \sim T R$ up
to a function of the angle on $S^5$. The case where
\nref{asymptvel} is zero corresponds to the case where the quark
and anti-quark have opposite scalar charges. In this case the
configuration is BPS
  and there is a family of
  static string solutions found in \cite{zachetal}.

A third situation that can be qualitatively modelled by
\nref{mends} is the one considered in \cite{matt}.
They introduce a heavy quark anti-quark pair separated by some distance
$L$ and add angular momentum $J$ along an $SO(2) \subset
SO(6)$. This corresponds to binding $J$ adjoint scalars to the heavy quark
anti-quark pair. If $J$ is very small, this corresponds to exciting
some small transverse fluctuations of the string
described around \nref{transos}. When
$J$ is large we can model the situation by a heavy particle at the tip, of
mass $ m \sim J/R$, up to a numerical coefficient which depends on
the details of the shape of the deformed string.
The bound states disappear for $m \sim T
R$, i.e., $J \sim T R^2 \sim \sqrt{g^2 N}$ (the precise numerical
coefficient was computed in \cite{matt}). Note that for weak
coupling there are no bound states.

Although the ladder approximation is obviously inadequate for
large `t Hooft coupling, it is encouraging that its strong
coupling limit is qualitatively similar to the exact AdS/CFT
results. The lowest energy eigenvalue has the $-\sqrt{\lambda}/L$
coupling dependence in both cases although the dimensionless
coefficients disagree for obvious reasons \cite{ericksonsz}.
Furthermore, in both cases the gap between the lowest energy state and
the continuum is populated with discrete levels, and the level
spacing of the near threshold states matches exactly at strong
coupling with strings containing one long fold. The big
qualitative difference remaining is a discrepancy in the density
of states far below the threshold, near the ground state. In the
AdS/CFT case the levels are those of a string, namely those of an
infinite number of oscillators with frequencies $\omega_n$,
$n=1,2\cdots$. In contrast, the sum of ladder diagrams predicts
discrete levels of a single harmonic oscillator. Presumably,
including more complicated planar diagrams, which include, for
example, long chains of gluons with near neighbor interactions
\cite{greensitet}, will remove this discrepancy.

Finally, we ask if the transition discussed above could also
arise in QCD. In a naive ladder diagram approximation,
we only have the gauge field exchange, but no scalar
exchange. We then seem to find that the new bound states appear at
 \bea \label{alphaqcd}
 {N\alpha_s\over\pi}>{1\over2},\quad {\rm for~QCD};\qquad
 {N\alpha_s\over\pi}>{1\over4},\quad {\rm for~{\cal N}=4}
 \eea For
$N=3$ this gives $\alpha^c_s=\pi/6\approx0.5$. This is not an
exact result in QCD, but rather an order of magnitude estimate.
Note that a flux tube in a pure glue theory without dynamical quarks
 cannot exhibit a sudden
appearance of states: because of confinement, the spectrum is completely
discrete, without a threshold. Lattice
studies indicate some crossover of gluonic excitation
levels between short string and long string regimes, at string
length around $1$ fm \cite{Kuti}. However, there is no
evidence for states with a $-1/L$ behavior of energy, other than the ground
state (see fig. 2 of \cite{Kuti}).
Thus, it appears that in QCD the effects of
asymptotic freedom do
not allow us to see the particular transition we discussed in this
article. Perhaps a simple qualitative way to explain this is to
note that the coefficient of the $1/r$ potential remains small
until the confining potential, linear in $r$, takes over.

\section{Strings with many folds}

\begin{figure}[ht]
\begin{center}
\includegraphics{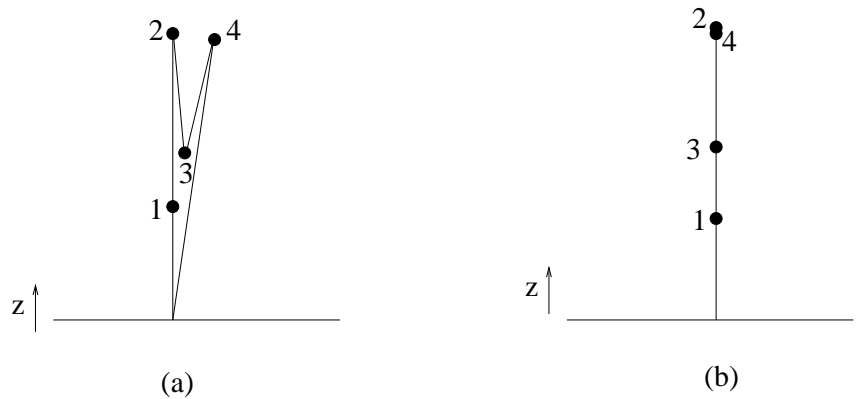}
\caption{ In (a)  we show a configuration of a string stretched in
the $z$ direction with pieces along the string carrying momentum
along the $z$ direction. To facilitate visualization we have
separated the strings in a transverse dimension, but we should
think of all of the pieces as lying on top of each other as in
(b).  } \label{folded}
\end{center}
\end{figure}

In this section we consider the dynamics of strings with many
folds. We consider the strict $L=0$ limit. In order to do this we
note that the whole motion of the string is taking place in an
$AdS_2$ subspace of $AdS_5$ parametrized by the coordinates
$t,~z$. Actually, the problem has a natural $SO(2,1) \times SO(3)$
symmetry. As in \cite{kapustin} it is convenient to think of the
gauge theory as defined on $AdS_2 \times S^2$. This is related by
a simple Weyl transformation to $R^{1,3}$. In this case it is
useful to parametrize $AdS_5$ as \be ds^2 = d\rho^2 + \cosh^2 \rho
ds^2_{AdS_2} + \sinh^2 \rho ds^2_{S^2} \ee The simplest BPS Wilson
loop  associated to a single quark insertion  corresponds to a
string worldsheet located at $\rho=0$. The folded string
configuration we want to consider is also at $\rho=0$. In a
semiclassical approximation, we can consider $n$ massless
particles that are joined by stretched strings \cite{folded}. Of
course, the ``massless particles'' are the parts of the string
that are not stretched and are carrying momentum. The potential
energy due to a stretched string segment going between two
particles  is \be \label{potpair} V = \alpha \left| { 1 \over z_1}
- { 1 \over z_2} \right| ~,~~~~~~\alpha = R^2 T = { \sqrt{g^2 N}
\over 2 \pi} \ee We now consider a quark at the boundary at
$z_0=0$ and a set of $n$ massless particles joined by strings and
ending at the boundary at $z_{n+1}=0$. \be \label{hampr}
 H = \sum_{i=1}^n |p_i| +
\sum_{i=0}^n V(z_i,z_{i+1})
\ee
It is interesting to compare this Hamiltonian with the naive Hamiltonian that we would
get for $n$ massless gluons that are in spherically symmetric wavefunctions and
interacting via Coulomb exchange at weak coupling, with the identification $r_i \sim z_i$.
 This would give a Hamiltonian precisely of the
same form, except that the coefficient $\alpha$ in \nref{potpair}
would be different, it would be $\alpha_{weak} = { g^2 N/ 4
\pi } $. Note that for $\alpha \ll 1$ a Hamiltonian such as
\nref{hampr} would not be a good classical approximation to the
quantum problem, while it is a good approximation if $\alpha \gg
1$. Note that we can think of the problem as follows. After a
Kaluza Klein reduction on the $S^2$, the resulting theory is
roughly like a two dimensional QCD theory with adjoint matter on
$AdS_2$. It is amusing that the ordinary Coulomb potential looks
confining in $AdS_2$. The Hamiltonian \nref{hampr} is reminiscent
of that found in the 2-d adjoint QCD \cite{Dalley}. To be more
precise, we have to include processes where a pair of adjacent
folds is created or destroyed. Such particle number changing
processes are also present in the adjoint QCD.

\section{Wilson lines in the adjoint representation}

The $L\to 0$ limit of the quark anti-quark potential is also
relevant for the analysis of the Wilson line in the adjoint
representation. 
This relation to the quark and anti-quark
Wilson lines shows that the adjoint Wilson line is not locally
BPS saturated.
 What we find here is that this Wilson line
operator has rather peculiar
properties at strong
coupling.

At weak coupling nothing dramatic happens.
There are of course the usual UV divergences from self-energy
corrections to the adjoint external source represented by the
Wilson line, but these can always be renormalized.
The divergence can be absorbed into a renormalization of the mass
of the external adjoint source. After subtracting this divergence the
spectrum of excitations, at weak coupling, is bounded from below. Since
the external adjoint exerts an attractive force on the dynamical
gluons, we need to use the uncertainty principle argument discussed
in section 2.1 in order to show that gluons will not lead to very negative
energy bound states at weak coupling.

On the other hand, according to the AdS/CFT correspondence,
at large $N$ and
strong `t Hooft coupling the adjoint Wilson
line introduces a folded string
as shown in figure \ref{stretched}(c). Such a string has infinite
energy since it stretches all the way to the boundary. This is not
surprising; it also happens for the Wilson loop in the
fundamental. This is a UV divergence which can be subtracted as in
\cite{maldacenaqqbar}.
Once we subtract this infinite part, we can further decrease the
energy by moving the tip of the fold towards the boundary. So,
after this subtraction, the Hamiltonian in this sector is unbounded
below, and any physical computation will involve a scattering
amplitude where the fold comes in from the boundary and goes back
to the boundary. Note that this feature of the Hamiltonian has real physical
meaning and it is not due to a wrong renormalization procedure.
This is the AdS dual of the ``fall to the center'' screening
of the external adjoint source
by a dynamical gluon when the coupling is sufficiently large.
The situation is similar to that encountered in the
two-dimensional string theory, dual to large $N$ matrix quantum
mechanics \cite{longstrings}.\footnote{In this case we can do the
computation on both sides of the duality and, after
renormalization,  we find a Hamiltonian that is unbounded below on
both sides.} This transition is related to the fact that the
linearized wave equation for a gluon in the field of an adjoint
source displays a ``fall to the center'' instability
\cite{mandula}.\footnote{We expect that, once we take into account
the fields created by the gluon, sources in the fundamental
representation will not cause a fall to the center instability for
a gluon.}
In other words, we
can interpret the tip of the folded string in figure
\ref{stretched}(c) as a gluon or group of gluons in an S-wave  and
the straight string pieces as coming from the Coulomb interaction.
The fact that the tip falls all the way to the boundary is then
related to the gluon falling to the center. This peculiar behavior
is also related to the fact that ``quarkonium"  is very small in
theories with a gravity dual \cite{adj,matt}.

As the coupling is reduced in the AdS description,
the discussion below (\ref{enerco}) shows that, below some critical coupling
the tip of the fold falls towards the horizon rather than towards
the boundary. This agrees with the absence of the
``fall to the center'' instability at weak coupling.
In real world QCD, since the coupling decreases
at short
distances, there is no ``fall to the center" instability; see the
discussion after \nref{alphaqcd}.

\section{Correlation functions of Wilson loops and
local operators }

In this section\footnote{ This section was written in response to
a question raised by E. Shuryak.}
 we will consider local vertex operators ${\cal O}$
and we will compute the expectation values of these vertex
operators in the presence of the Wilson loop corresponding to a
static quark and antiquark separated by distance $L$. Let us first
consider an operator ${\cal O}_J = \Tr[\Phi^J] +\cdots $, where
${\cal O}_J$ is a BPS operator of dimension $\Delta =J$ in the
${\cal N}=4$ theory. The first guess for the weak coupling scaling
of such correlation functions is \be \label{naiveans}
 { \langle W {\cal O}(x) \rangle \over \langle W \rangle }
\sim   { 1 \over x^{\Delta } }
 \ee
at large $x$, where $x$ is the distance of the operator insertion
from the quark-antiquark pair, $x \gg L$.

The strong coupling answer may be computed using the AdS/CFT
duality as in \cite{corrado}
 \be \label{strongcu}
 { \langle W {\cal O}(x) \rangle \over \langle W \rangle } \
 \sim  { \sqrt{g^2 N} \over N}
 { L^{\Delta -1} \over x^{2 \Delta -1 } }
 \ee
The factor $\sqrt{g^2 N}$ originates from $R^2/\alpha'$
in the string action. This
result holds for any operator dual to a supergravity field,
up to a numerical factor, which
could be zero for some operators due to symmetry reasons.

Note that the fall-off as a function of $x$ in \nref{strongcu} is
faster for $\Delta>1$. For specific operators, such as ${\cal O}'_4 =
\Tr[F_{\mu\nu}^2]$, the angular dependence also differs from the
naive weak coupling expectation that the expectation value of $\Tr E^2$
is given by the square of the dipole electric field $\sim  L^2(1 +
3 \cos^2 \theta ) /x^6$. The latter answer would be obtained in
the $U(1)$ gauge theory, but the large $N$ theory is very
different from it due to the planarity restriction. In fact, we
will show that the correct {\it planar} weak coupling answer has
the same dependence on $x$ as the strong coupling result
\nref{strongcu}.

\begin{figure}[ht]
\begin{center}
\includegraphics{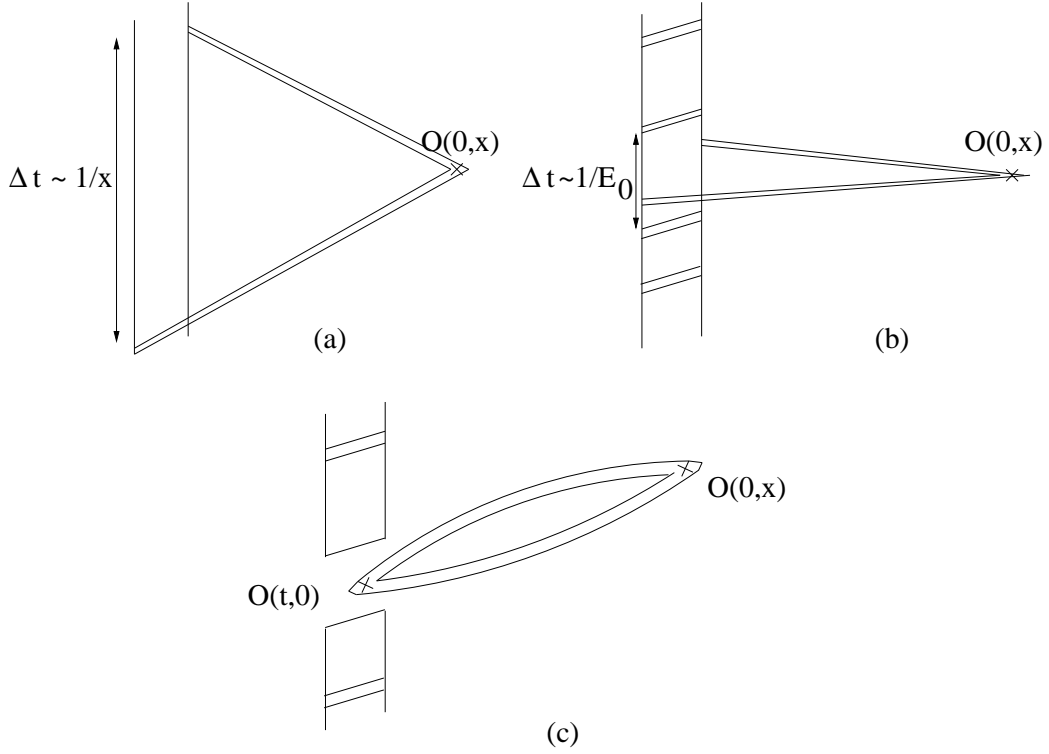}
\caption{ In (a) we see the diagrams contributing to the naive
answer. It contains diagrams where the propagator endpoints are
separated by a distance of the order of $\Delta t \sim 1/x$. In
(b) we take into account the diagrams contributing to Coulomb
energy between the quark and antiquark. Planarity plus energetic
considerations effectively restrict the endpoints of the
propagators to be within $\Delta t \sim 1/E_0$.
  In (c) we replace the square in the center of (b) by the
  insertion
 of a series of operators which is also integrated along $t$,
 the direction of the original
 contour.   }
\label{localopetwo}
\end{center}
\end{figure}

The crucial difference between the $U(1)$ and
the planar calculations is the following. For the
$U(1)$ calculation
the Coulomb exchanges between the charges play no role.
We find field
propagators connecting the operator insertion and the quark lines,
 see figure \ref{localopetwo}(a). The $U(1)$ answer,
such as \nref{naiveans}, arises from integrating the end-points of
such propagators all the way along the time direction. If $x$ is
large, then the result \nref{naiveans} has contributions from the
end points of the propagators on the quark lines separated by
distances $\Delta t \sim x $ from each other.

At large $N$, the situation is different because the gluon
exchanges between the quark and anti-quark keep changing their
colors, but the pair of gluon fields emanating from $O_4$, for
example, has correlated colors. For this reason, the positions of
these lines cannot be integrated along the entire $t$-direction.
The correct answer takes into account the fact that these gluon
propagators, if placed at very different times along the quark and
anti-quark lines, would remove the Coulomb attractive energy, much
in the same way as what we discussed for gluon radiation
in Section 2. At
weak coupling, the gluons that contribute to the Coulomb
interaction energy are separated by a distance of the order of
$\Delta t \sim 1/E_0 \sim L/(g^2 N)$. Thus, the endpoints of the
gluons connecting the operators to the quark lines should all end
within a distance of the order of $\Delta t$ of each other. If the
gluon endpoint separations are bigger than $\Delta t$, we should
get an exponential suppression due to the loss of the Coulomb
binding energy. Thus, for distances $ L/(g^2 N) \gg x \gg L $ the
large $N$ weak `t Hooft coupling answer agrees qualitatively with
the $U(1)$ intuition. But for $x\gg L/(g^2 N)$ we find the same
scaling as in (\ref{strongcu}).

Our conclusion is that, for distances $x \gg L/(g^2 N)$, we can
replace the endpoints of all propagators on the quark lines
(suitably integrated) by a series of local color-singlet
operators, see figure \ref{localopetwo}(c). Similarly, we can have
graphs where the pair of gluons emanating from ${\cal O}$ is
inserted on the same gluon line in the ladder graph. We can think
of all the relevant graphs as propagation of a color-singlet state
from the location of ${\cal O}$ to an insertion somewhere on the
QCD string worldsheet. Such a description through color singlet
diagrams is similar to what one does on the string side of the
AdS/CFT correspondence. We emphasize that for sufficiently large
$y$ this picture is valid at both weak and strong coupling of the
large $N$ gauge theory.

At this point we can replace a narrow Wilson loop by a series of
color singlet operator insertions. The external operator ${\cal
O}$ will give a non-trivial two point function only if ${\cal O}$
itself appears in this OPE. Let us consider first the operators
${\cal O}_J \sim  \Tr[\Phi^J]$.   Normalizing the two point
functions of ${\cal O}_J$ to $1$, the relevant term in the Wilson
loop expansion will have the form \be \label{constdef}
 c_w \int dt   {\cal O}_J ~,~~~~~~~c_w \sim
 {  L^{J -1} \over (g^2 N)^{J -1} } { ( g^2 N)^{J/2} } { 1 \over N}
\ee where the integral is along the Euclidean time $t$ and the
operators are inserted at $x_1 =0$ (recall that the quark and
antiquark are located at $x_1 = \pm L/2$ respectively). The first
factor in $c_w$ \nref{constdef} comes from  integrating the
endpoints of the propagators over a restricted range, see figure
\ref{localopetwo}(b). The second comes from the normalization of
the operator and powers of $N$ that can be counted by looking at
the difference between figures \ref{localopetwo}(b)
\ref{localopetwo}(c).

Then the correct planar answer at weak `t Hooft coupling is \be
\label{corrwe} { \langle W {\cal O}(x) \rangle \over \langle W
\rangle } \sim  c_w \int dt
  \langle {\cal O}_J(t,0) {\cal O}_J(0, x)  \rangle =   c_w \int dt
  {1 \over ( t^2 + x^2)^{\Delta} } \sim c_w \,  { 1 \over x^{2 \Delta -1} }
\ee where $c_w$ is the constant appearing in \nref{constdef}. This
result is correct for distances $x \gg L/(g^2 N)$. Remarkably,
this has the same structure as the strong coupling answer
\nref{strongcu}, except that the power of $g^2 N$ is different. We
can
 reproduce the dependence of \nref{strongcu}
on the 't Hooft coupling if we assume
 that the propagator endpoints
are localized within a distance of the order of
$\Delta t \sim L/\sqrt{g^2 N}$, which
is indeed the distance at which the strong coupling
Coulomb potential becomes important \cite{Shuryak}.

In particular, this reproduces the result of \cite{callang} for
the operator
 $\langle W \Tr [F_{\mu\nu}^2] \rangle $, $\sim L^3/x^7$.
At weak coupling one might naively apply the $U(1)$ intuition
suggesting the behavior as a square of a dipole field $  L^2(1 + 3
\cos^2 \theta )/x^6$ where $\theta$ is the angle between $\vec x$
and the line between the quark-antiquark pair. Our arguments show
that the correct weak coupling result at large $N$ and large
enough $x$ has the same isotropic behavior as the strong coupling
result of \cite{callang}. These arguments are also valid for the
stress tensor. So the expectation value for the energy $T_{00}$
also behaves as $L^3/x^7$ both at weak and strong coupling.

Finally, note that there may be $1/N$ suppressed contributions
coming
from non-planar diagrams, which decay more slowly than
\nref{corrwe} at long distances.
 Thus for distances $x$
which are parametrically large in $N$ these contributions will
dominate. Similar contributions could be present at strong
coupling also. Presumably they come from loop diagrams in $AdS$,
but we have not found their precise $x$ dependence.

\section{Discussion}

One of the lessons from this article is that, already at weak
coupling, large $N$ effects produce some results that
are in rough qualitative
agreement with the strong coupling results provided by
the AdS/CFT correspondence. One
example is the existence of a gap for excitations of the flux tube
connecting a heavy quark anti-quark pair. The second is the large
$x$ behavior of one point functions.

On the other hand, some features appear only at strong coupling;
for example, the existence of an infinite number of bound state
excitations of the flux tube. In this article we studied excited
states of this flux tube near the  ``ionization" threshold. These
states correspond to long folded strings in $AdS$. The density of
states, computed with the ladder model, matched precisely the
strong coupling answer based on a string with a single fold.
Furthermore, the dynamics of a string with many folds resembles
quite closely the expected dynamics of S-wave gluons in the gauge
theory.

We might ask whether the transition we discussed here is evidence
for a quantum phase transition in ${\cal N}=4$ Yang Mills theory
as a function of $g^2 N$. We believe that the answer is no. In
fact, perfectly analytic systems like ordinary quantum mechanical
systems can display transitions in the structure of the spectrum
without there being a phase
transition. In fact, we also found that
some of the strong coupling features, such as the long distance
behavior of one point functions, are also present at weak
coupling.

Finally, we mention that QCD flux tube excitations may
play a role
in sufficiently excited mesons (although, as mentioned above,
these excitations occur in the linear confining regime
of the flux tube). For example,
some of the new mesons recently observed
above the $D\bar D$ threshold in the ${\bf c \bar
c }$ system (see \cite{quigg} for a review)
could be gluonic excitations of charmonium. In other words,
perhaps some of
these new mesons can be interpreted as hybrid states containing
extra gluons.

\underline{Acknowledgments:}

We would like to thank C. Callan and E. Shuryak for asking us
the questions which prompted us to think about these problems.
C.B.T. would like to thank  R. Brower and C.-I. Tan for
helpful discussions.
This work was supported by the National Science Foundation under
Grant No. PHY-0243680, and the U.S.\ Department of Energy under
Grants No. DE-FG02-97ER-41029 and
 DE-FG02-90ER40542.
Any opinions, findings, and conclusions or recommendations expressed in
this material are those of the authors and do not necessarily reflect
the views of the National Science Foundation.

\appendix
\section{Solutions for strings ending at different points on $S^5$}

We present a  time dependent solution which describes strings
sitting at a spatial point on the boundary but ending at two
points on $S^5$ separated by some angle $\Delta \theta$.
 We have a string on the
space \be\label{metrrel}
 ds^2 = { - dt^2 + d z^2 \over z^2 } + d\theta^2
\ee
  We write the Nambu
action in variables where $t = \tau$ and $\theta = \sigma$
 \be\label{actth} \int dt d\theta \sqrt{ { 1 - \dot z^2 \over z^2} + {
  {z'}^2 \over z^4 } }
  \ee
  where the prime denotes the derivative with respect to $\theta$.
  We now guess a solution with a simple time dependence of
   the form $z = \tau f(\theta)$.
   The equation for
  $f$ can be integrated once and we find
  \be\label{conservq}
  { 1 - f^2 \over \sqrt{ (1-f^2)f^2 + {f'}^2 } } = { \sqrt{1
  -f_0^2 } \over f_0 } ~,~~~~{\rm or}   ~~~{ df \sqrt{1 -f_0^2} \over
  \sqrt{ (1-f^2)(f_0^2 - f^2) } } = d\theta
   \ee
  where $ 0<f_0<1$ is the point where $f'=0$, which we take to be at
  $\theta =0$.   Notice that $f_0 = { \partial_\tau z}|_{\theta
  =0}$ is the speed of the motion of the tip. The equation \nref{conservq} determines
  $f(\theta)$.
 We find
\be
 {\Delta \theta \over 2} =  \sqrt{ 1
 - f_0^2} \int_0^{f_0} { d f \over \sqrt{ (1 - f^2)(f_0^2 - f^2)} }= \sqrt{ 1
 - f_0^2}  K[ f_0]
 \ee
 where $K$ is an elliptic function (see \cite{gradsh}).
 This equation
 gives us the velocity for the motion of the tip in terms
 of $\Delta \theta$. For $\Delta \theta =\pi$ the velocity goes to
 zero.

\end{document}